\documentclass[12pt]{article}
\usepackage[dvips]{graphicx}
\usepackage{epsfig}
\usepackage{amsmath,amssymb,amsbsy}
\usepackage{amsfonts}
\begin{document}
\thispagestyle{empty}
\begin{center}

{\Large\bf{
Statistical description of the proton spin \\
\vskip 0.5cm
with a large gluon helicity distribution}}
\vskip1.4cm
{\bf Claude Bourrely}
\vskip 0.3cm
Aix-Marseille Universit\'e, D\'epartement de Physique, \\
Facult\'e des Sciences site de Luminy, 13288 Marseille, Cedex 09, France\\
\vskip 0.5cm
{\bf Jacques Soffer}
\vskip 0.3cm
Physics Department, Temple University\\
1835 N, 12th Street, Philadelphia, PA 19122-6082, USA
\vskip 0.5cm
{\bf Abstract}\end{center}

The quantum statistical parton distributions approach proposed more than one decade ago is revisited by considering a larger set of recent and accurate
Deep Inelastic Scattering (DIS) experimental results. It enables us to improve the description of the data by means of a new determination of the parton distributions. We will see that a large gluon polarization emerges, giving a significant contribution to the proton spin.

\vskip 0.5cm

\noindent {\it Key words}: Gluon polarization; Proton spin; Statistical distributions\\

\noindent PACS numbers: 12.40.Ee, 13.60.Hb, 13.88.+e, 14.70.Dj
\vskip 0.5cm

\newpage
\section{Introduction}
 Several years ago a new set of parton distribution functions (PDF) was constructed in the
framework of a statistical approach of the nucleon \cite{bbs1}. For quarks
(antiquarks), the building blocks are the helicity dependent distributions
$q^{\pm}(x)$ ($\bar q^{\pm}(x)$). This allows to describe simultaneously the
unpolarized distributions $q(x)= q^{+}(x)+q^{-}(x)$ and the helicity
distributions $\Delta q(x) = q^{+}(x)-q^{-}(x)$ (similarly for antiquarks). At
the initial energy scale, these distributions
are given by the sum of two terms, a quasi Fermi-Dirac function and a helicity
independent diffractive
contribution. The flavor asymmetry for the light sea, {\it i.e.} $\bar d (x) > \bar
u (x)$, observed in the data is built in. This is simply understood in terms
of the Pauli exclusion principle, based on the fact that the proton contains
two up-quarks and only one down-quark. The chiral properties of QCD lead to
strong relations between $q(x)$ and $\bar q (x)$.
For example, it is found that the well estalished result $\Delta u (x)>0 $\
implies $\Delta
\bar u (x)>0$ and similarly $\Delta d (x)<0$ leads to $\Delta \bar d (x)<0$. 
This earlier prediction was confirmed by recent data. In addition we found the approximate equality of the flavor asymmetries, namely $\bar d(x) - \bar u(x) \sim \Delta \bar u(x) - \Delta \bar d(x)$. Concerning
 the gluon, the unpolarized distribution $G(x,Q_0^2)$ is given in
terms of a quasi Bose-Einstein function, with only {\it one free parameter},
and for simplicity, we were assuming zero gluon polarization, {\it i.e.} $\Delta
G(x,Q_0^2)=0$, at the initial energy scale $Q_0^2$. As we will see below, the new analysis of a larger set of recent accurate DIS data, has enforced us to
to give up this assumption. It leads to an unexpected large gluon helicity distribution and this is the major point, which is emphasized in this letter. 
In our previous analysis all unpolarized and
helicity light quark distributions were depending upon {\it eight
free parameters}, which were determined in 2002 (see Ref.~\cite{bbs1}), from a
next-to-leading (NLO) fit of a small set of accurate DIS data. Concerning
the strange quarks and antiquarks distributions, the statistical approach was
applied using slightly different expressions, with four additional parameters \cite{bbs2}. Since the first determination of the free parameters, new tests against experimental (unpolarized and
polarized) data turned out to be very satisfactory, in particular in hadronic
reactions, as reported in Refs.~\cite{bbs3,bbs4,bbsW}.\\
In this letter, after a brief review of the statistical approach, we will only give some elements of the new determination of the parton distributions,
but we will focus on the gluon helicity distribution, a fundamental contribution to the proton spin.

\section{Basic review on the statistical approach}
Let us now recall the main features of the statistical approach for building up the PDF, as oppose
to the standard polynomial type
parametrizations of the PDF, based on Regge theory at low $x$ and on counting
rules at large $x$.
The fermion distributions are given by the sum of two terms,
a quasi Fermi-Dirac function and a helicity independent diffractive
contribution:
\begin{equation}
xq^h(x,Q^2_0)=
\frac{A_{q}X^h_{0q}x^{b_q}}{\exp [(x-X^h_{0q})/\bar{x}]+1}+
\frac{\tilde{A}_{q}x^{\tilde{b}_{q}}}{\exp(x/\bar{x})+1}~,
\label{eq1}
\end{equation}
\begin{equation}
x\bar{q}^h(x,Q^2_0)=
\frac{{\bar A_{q}}(X^{-h}_{0q})^{-1}x^{b_{\bar q}}}{\exp [(x+X^{-h}_{0q})/\bar{x}]+1}+
\frac{\tilde{A}_{q}x^{\tilde{b}_{q}}}{\exp(x/\bar{x})+1}~,
\label{eq2}
\end{equation}
at the input energy scale $Q_0^2=1 \mbox{GeV}^2$. We note that the diffractive term is absent in the quark helicity distribution $\Delta q$ and in the quark valence contribution $q - \bar q$.\\
In Eqs.~(\ref{eq1},\ref{eq2}) the multiplicative factors $X^{h}_{0q}$ and
$(X^{-h}_{0q})^{-1}$ in
the numerators of the non-diffractive parts of the $q$'s and $\bar{q}$'s
distributions, imply a modification
of the quantum statistical form, we were led to propose in order to agree with
experimental data. The presence of these multiplicative factors was justified
in our earlier attempt to generate the transverse momentum dependence (TMD)
\cite{bbs5}, which was revisited recently \cite{bbs6}.
The parameter $\bar{x}$ plays the role of a {\it universal temperature}
and $X^{\pm}_{0q}$ are the two {\it thermodynamical potentials} of the quark
$q$, with helicity $h=\pm$. Notice the change of sign of the potentials
and helicity for the antiquarks \footnote{~At variance with statistical mechanics where the distributions are expressed in terms of the energy, here one uses
 $x$ which is clearly the natural variable entering in all the sum rules of the parton model.}.\\
For a given flavor $q$ the corresponding quark and antiquark distributions involve {\it eight} free parameters: $X^{\pm}_{0q}$, $A_q$, $\bar {A}_q$, $\tilde {A}_q$, $b_q$, $\bar {b}_q$ and $\tilde {b}_q$. It reduces to $\it seven$ since one of them is fixed by the valence sum rule, $\int (q(x) - \bar {q}(x))dx = N_q$, where $N_q = 2, 1, 0 ~~\mbox{for}~~ u, d, s$, respectively.  
For the light quarks $q=u,d$,  the total number of free parameters is reduced to $\it eight$ by taking, as in Ref. \cite{bbs1}, $A_u=A_d$, $\bar {A}_u = \bar {A}_d$, $\tilde {A}_u = \tilde {A}_d$, $b_u = b_d$, $\bar {b}_u = \bar {b}_d$ and $\tilde {b}_u = \tilde {b}_d$. For the strange quark and antiquark distributions, the simple choice made in Ref. \cite{bbs1}
was improved in Ref. \cite{bbs2}, but here they are expressed in terms of $\it seven$ free parameters.\\
For the gluons we consider the black-body inspired expression
\begin{equation}
xG(x,Q^2_0)=
\frac{A_Gx^{b_G}}{\exp(x/\bar{x})-1}~,
\label{eq3}
\end{equation}
a quasi Bose-Einstein function, with $b_G$ being the only free parameter, since $A_G$ is determined by the momentum sum rule.
In our earlier works \cite{bbs1,bbs4}, we were assuming that, at the input energy scale, the polarized gluon,
distribution vanishes, so
\begin{equation}
x\Delta G(x,Q^2_0)=0~.
\label{eq4}
\end{equation}
However in our recent analysis of a much larger set of very accurate unpolarized and polarized DIS data, we must give up this simplifying assumption. We are now taking
\begin{equation}
\Delta G(x,Q^2_0)=P(x)G(x,Q^2_0)~,
\label{eq5}
\end{equation}
where $P(x)$ is expressed in terms of $\it four$ free parameters, as follows
\begin{equation}
 P(x) = \bar A_{G}x^{\bar b_{G}}/(c_G + x^{-d_G})~.
\label{eq6}
\end{equation}
We must have $|P(x)| \leq 1$, to insure that positivity is satisfied.\\
To summarize the new determination of the PDF involves a total of {\it twenty one} free parameters: in addition to the temperature $\bar x$ and the exponent $b_G$ of the gluon distribution, we have {\it eight} free parameters for the light quarks $(u,d)$, {\it seven} free parameters for the strange quarks and {\it four} free parameters for the gluon helicity distribution. These parameters were determined from a next-to leading order (NLO) fit of 
a large set of accurate DIS data, (the unpolarized structure functions $F_2^{p,n,d}(x,Q^2)$, the polarized structure functions $g_1^{p,n,d}(x,Q^2)$, the structure function $xF_3^{\nu N}(x,Q^2)$ in $\nu N$ DIS, etc...) a total of 2140 experimental points with an average $\chi^2/pt$ of 1.33. Although the full details of these new results will be presented in a forthcoming paper \cite{bbs14}, we just want to make a general remark. By comparing with the results of 2002 \cite{bbs1}, we have observed a remarquable stability of some important parameters, the light quarks potentials $X_{0u}^{\pm}$ and  $X_{0d}^{\pm}$, whose numerical values are almost unchanged. The new temperature is slightly lower. As a result the main features of the new light quark and antiquark distributions are only hardly modified, which is not surprizing, since our 2002 PDF set has proven to have a rather good predictive power.\\
We now turn to the gluon helicity distribution which is the main purpose of this letter.

\section{The gluon helicity distribution and the proton spin}
\begin{figure}[htp]
\vspace*{-3.5ex}
\begin{center}
\includegraphics[width=14.0cm]{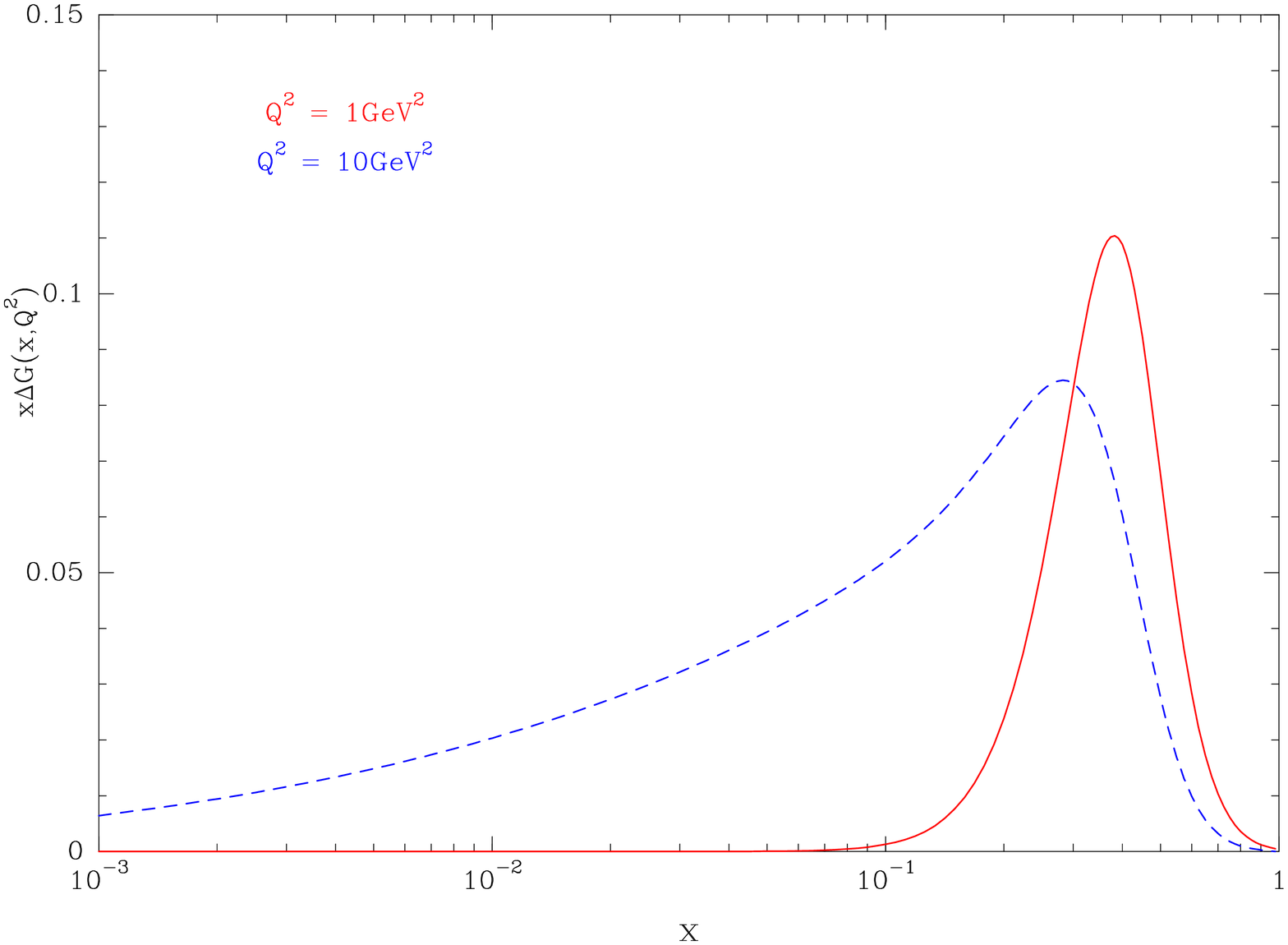}
\caption[*]{\baselineskip 1pt
 The gluon helicity distribution $x\Delta G(x,Q^2)$ versus $x$, for $Q^2=1\mbox{GeV}^2$ (solid curve) and $Q^2=10\mbox{GeV}^2$ (dashed curve) }
\label{gluon-hel}
\end{center}
\end{figure}

\begin{figure}[htp]
\vspace*{-3.5ex}
\begin{center}
\includegraphics[width=14.0cm]{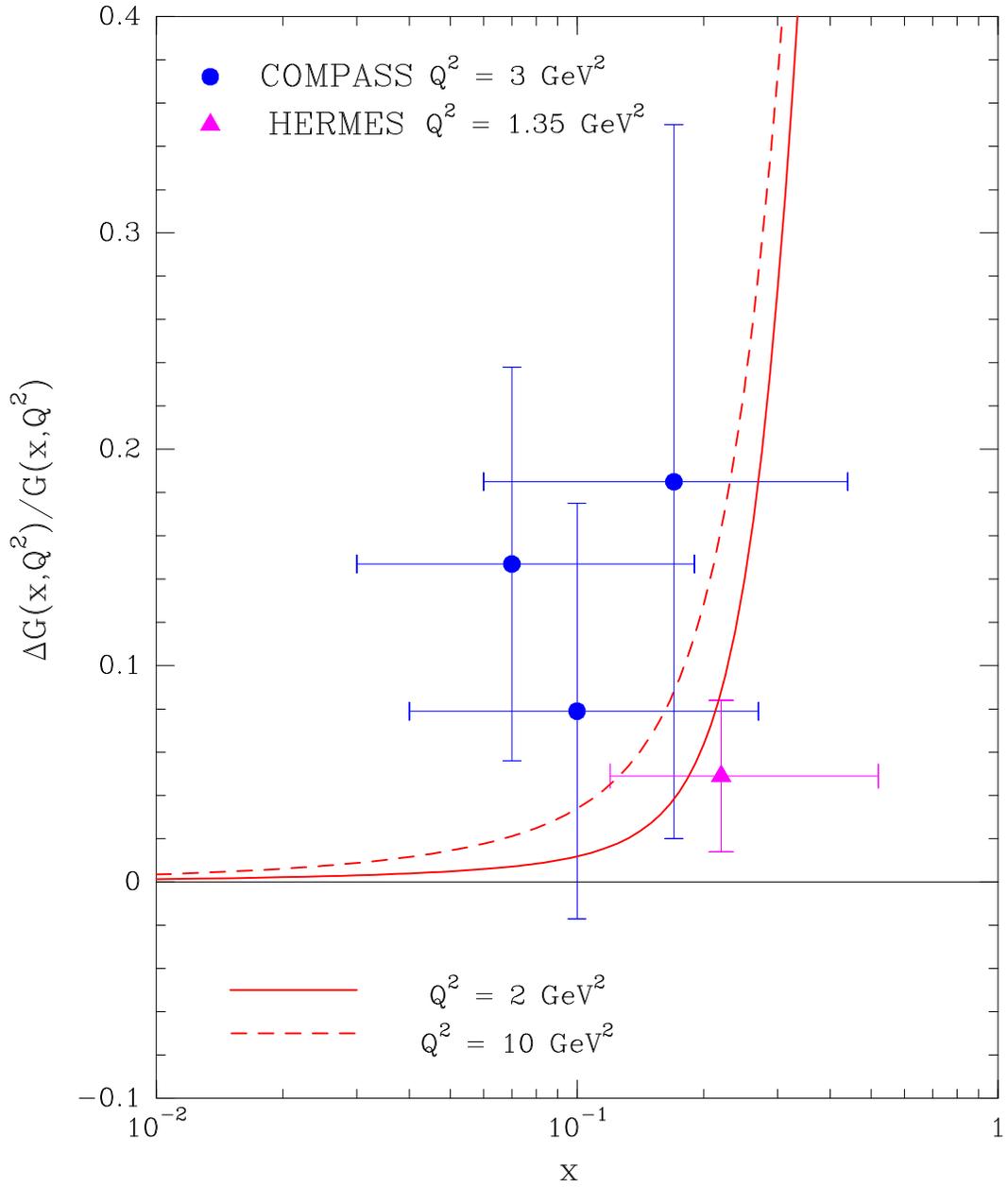}
\caption[*]{\baselineskip 1pt
$\Delta G(x,Q^2)/G(x,Q^2)$ versus $x$, for $Q^2=2\mbox{GeV}^2$ (solid curve) and $Q^2=10\mbox{GeV}^2$ (dashed curve). The data are from HERMES \cite{hermes} and COMPASS \cite{compass}.}
\label{deltagog}
\end{center}
\end{figure}

\begin{figure}[htp]
\vspace*{-3.5ex}
\begin{center}
\includegraphics[width=14.0cm]{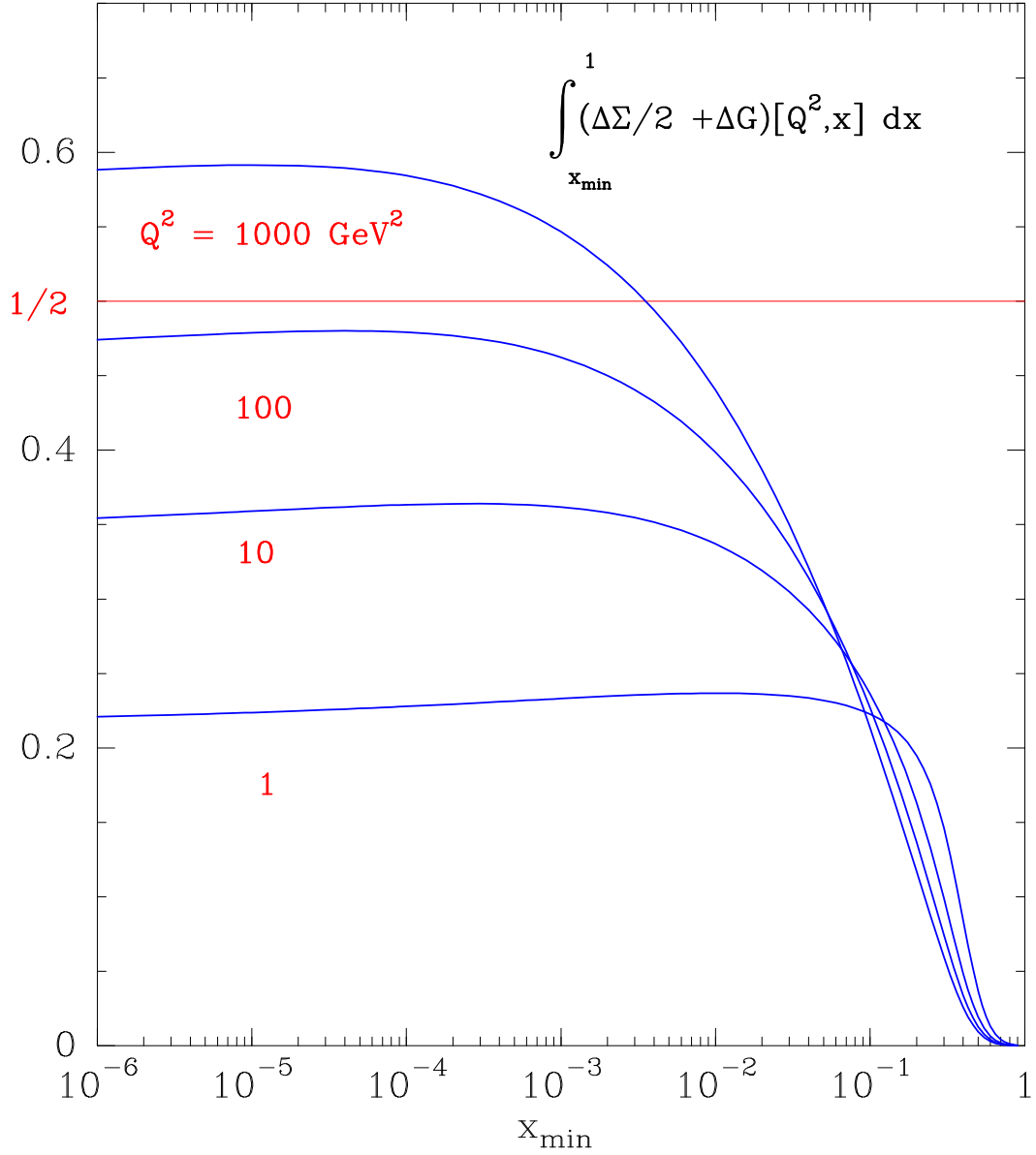}
\caption[*]{\baselineskip 1pt
 Total contribution of the quark, antiquark $\Delta \Sigma (x,Q^2)$ and gluon $\Delta G (x,Q^2)$ helicity densities, to the proton helicity sum rule versus the lower limit of the integral $x_{\mbox{min}}$, for different $Q^2$ values.}
\label{spinprot}
\end{center}
\end{figure}

\begin{figure}[htp]
\vspace*{-3.5ex}
\begin{center}
\includegraphics[width=14.0cm]{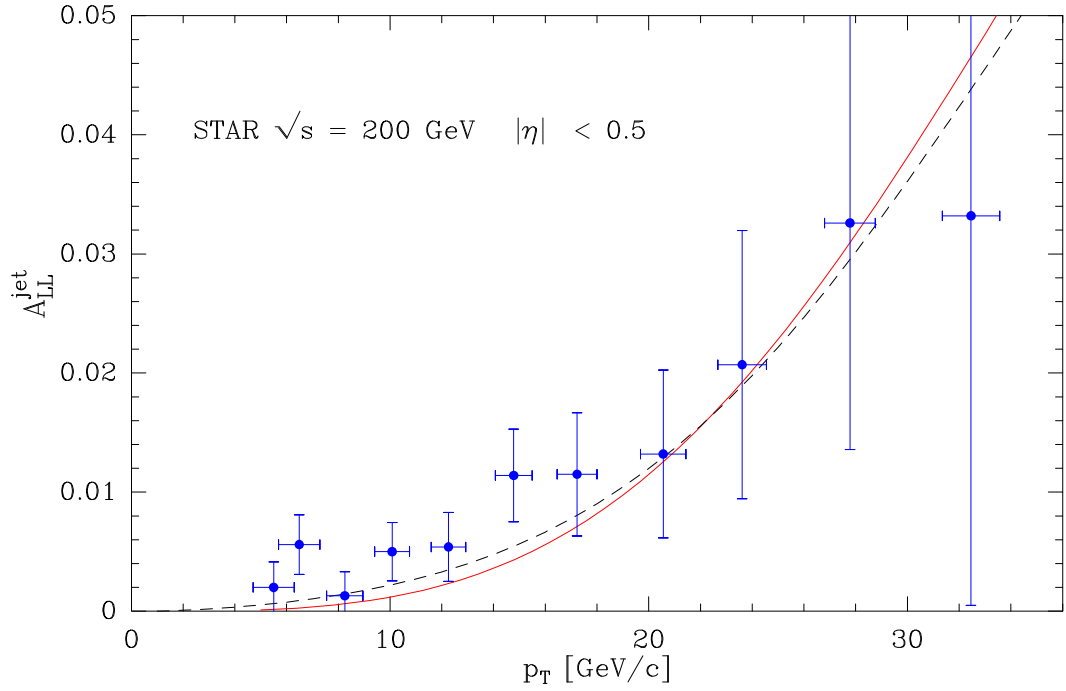}
\caption[*]{\baselineskip 1pt
{\it Solid curve}: Our predicted double-helicity asymmetry $A_{LL}^{jet}$ for jet production at BNL-RHIC in the near-forward rapidity region, versus $p_T$ and the data points from Ref. \cite{star}.\\
{\it Dashed curve}: The approximate expression Eq. (\ref{approx}) with k=1/7.}
\label{all}
\end{center}
\end{figure}

The gluon helicity distribution $\Delta G(x,Q^2)$ of the proton is a fundamental physical quantity for our understanding of the proton spin. Its integral
over $x$, $\Delta G(Q^2)$, may be interpreted, in the light-cone gauge $A^+=0$, as the gluon spin contribution to the proton spin \cite{LL}. If $\Delta \Sigma(Q^2)$ denotes the total sum of quark and antiquark helicity contributions and $L_{q,\bar q,G}(Q^2)$ are the quark, antiquark and gluon orbital angular momentum contributions, the proton helicity sum rule reads
\begin{equation}
1/2 =1/2 \Delta \Sigma(Q^2) + \Delta G(Q^2) + L_{q}(Q^2) + L_{\bar q}(Q^2) + L_G(Q^2) ~.
\label{eq7}
\end{equation}
In the above sum rule $\Delta \Sigma(Q^2)$ is certainly the contribution which is best known, with a typical value $\sim 0.3$ according to Refs. \cite{dssv1,others} and also from our own result. This contribution is therefore too small to satisfy the sum rule and it is crucial to find out how much the gluon 
contributes to the proton spin, a long standing problem.\\
The new determination of the PDF leads, in the gluon sector, to the following parameters:
\begin{equation}
A_G = 36.778,~b_G = 1.020,~\bar {A}_G = 26.887,~\bar {b}_G = 0.162,~c_G = 0.006, ~d_G = 6.072~,
\label{eq8}
\end{equation}
and the new temperature is $\bar x =0.090$. We display in Fig. \ref{gluon-hel} the gluon helicity distribution versus $x$ at the initial scale $Q_{0}^2 = 1\mbox{GeV}^2$ and $Q^2 = 10\mbox{GeV}^2$. At the initial scale it is sharply peaked around $x=0.4$, but this feature lessens after some evolution. We find that $P(x)= 0.731 x^{5.210}/(x^{6.072} + 0.006)$, which is such that $0<P(x)<1$ for $0<x<1$, so positivity is satisfied and in addition the gluon helicity
distribution remains positive.\\
We display $\Delta G(x,Q^2)/G(x,Q^2)$ in  Fig. \ref{deltagog} for two $Q^2$ values and some data points \cite{hermes,compass} suggesting that the gluon helicity distribution is positive indeed. In several of the available parametrisations this ratio goes to zero when $x=1$, but we observe that, since for example at the initial scale  $P(x=1)=0.726$,  this is not the case here.\\
Let us now examine the consequences of this new gluon helicity distribution, with a rather strong $Q^2$ dependence, on the proton helicity sum rule Eq. (\ref{eq7}). By considering only the quark, antiquark and gluon helicity densities, we display in Fig. \ref{spinprot} their contributions versus the lower limit of the integral $x_{\mbox{min}}$, for different $Q^2$ values. For increasing $Q^2$ they slowly saturate the sum rule, allowing a decreassing positive contribution to the orbital angular momentum, which definitely must change sign for $Q^2= 1000\mbox{GeV}^2$.\\
Finally, we would like to test our new gluon helicity distribution in a pure hadronic reaction. In a very recent paper, the STAR Collaboration
at BNL-RHIC has reported the observation, in one-jet inclusive production, of a non-vanishing positive double-helicity asymmetry $A_{LL}^{\it jet}$ for $5 \leq p_T \leq 30 \mbox{GeV}$, in the near-forward rapidity region \cite{star}. We show in Fig. \ref{all} our prediction \footnote{We are grateful to Prof. W. Vogelsang for providing us with the code to make this calculation} compared with these high-statistics data points and the agreement is very reasonable. There is a simple way to understand the trend of this double-helicity asymmetry. In this kinematic region, where the jet has a pseudo-rapidity $\eta$ close to zero and a moderate $p_T$, the dominant subprocess is $uG \to uG$, so we can write the following approximate expression
\begin{equation}
A_{LL}^{\it jet}= k \frac{\Delta G (x_T)}{G(x_T)}\cdot\frac{\Delta u (x_T)}{u(x_T)}~,
\label{approx}
\end{equation}
where $x_T = 2p_T/\sqrt{s}$ and k is a normalisation factor such that $0 \leq k \leq 1$. It exhibits the strong correlation of $A_{LL}^{\it jet}$ on the sign
and magnitude of $\Delta G$ and the validity of this approximation is clearly shown in Fig. \ref{all}.

 This new STAR data has been used recently by the DSSV Collaboration \cite{dssv2} to perform a new global polarized fit which leads them to extract also a rather large positive gluon helicty distribution.\\
Although we cannot yet firmly claim the discovery of a large positive gluon helicity distribution, giving a significant contribution to the proton spin, these new results are strongly suggesting that we may have reached a benchmark in our knowledge of the nucleon structure. Other independent processes sensitive to $\Delta G(x,Q^2)$ must be investigated and in particular in $pp$ collisions, the di-jet production at froward rapidity is now strongly considered
at BNL-RHIC.

\clearpage
\newpage

\clearpage
\newpage

\clearpage
\newpage

\clearpage
\newpage

\end{document}